\documentclass{article}

\PassOptionsToPackage{numbers}{natbib}
\usepackage[final]{nips_2017}

\usepackage[utf8]{inputenc} 
\usepackage[T1]{fontenc}    
\usepackage{hyperref}       
\usepackage{url}            
\usepackage{booktabs}       
\usepackage{amsfonts}       
\usepackage{nicefrac}       
\usepackage{microtype}      

\usepackage[pdftex]{graphicx}
\usepackage{listings}

\title{Synkhronos: a Multi-GPU Theano Extension for Data Parallelism}

\author{
  Adam Stooke \\
  University of California, Berkeley \\
  \texttt{adam.stooke@berkeley.edu} \\
  \And
  Pieter Abbeel \\
  University of California, Berkeley \\
  \texttt{pabbeel@cs.berkeley.edu}
}

\begin{document}

\maketitle

\begin{abstract}
  We present Synkhronos, an extension to Theano for multi-GPU computations leveraging data parallelism.  Our framework provides automated execution and synchronization across devices, allowing users to continue to write serial programs without risk of race conditions.  The NVIDIA Collective Communication Library is used for high-bandwidth inter-GPU communication.  Further enhancements to the Theano function interface include input slicing (with aggregation) and input indexing, which perform common data-parallel computation patterns efficiently.  One example use case is synchronous SGD, which has recently been shown to scale well for a growing set of deep learning problems.  When training ResNet-50, we achieve a near-linear speedup of 7.5x on an NVIDIA DGX-1 using 8 GPUs, relative to Theano-only code running a single GPU in isolation.  Yet Synkhronos remains general to any data-parallel computation programmable in Theano.  By implementing parallelism at the level of individual Theano functions, our framework uniquely addresses a niche between manual multi-device programming and prescribed multi-GPU training routines.  
\end{abstract}

\section{Introduction}
Theano \cite{Theano} is the classic auto-differentiation Python package.  It translates user's computation expressions into optimized code for fast CPU or GPU execution, and deep learning is among its most common uses.  Yet it lacks built-in support for multi-device parallel programming,\footnote{support for multi-GPU model parallelism is still experimental} a clear means to faster computing.  While significant speedups in multi-core CPU execution are possible,\footnote{In our experience with some BLAS routines (called by Theano), calling separate single-threaded routines on each core in a data-parallel fashion can out-perform the multi-threaded routine using all cores on the full dataset.} we focus on use of GPUs.  

\subsection{Other Frameworks}
More recent auto-differentiation frameworks do offer multi-GPU programming in Python, under data or model parallelism.  For example, in Tensorflow \cite{tensorflow2015-whitepaper}, perhaps the most similar to Theano, users can program ``multi-tower'' computations\footnote{\texttt{https://www.tensorflow.org/tutorials/using\_gpu}} with more or less arbitrary device placements for computations.  Other successful packages, such as PyTorch \cite{pytorch}, Chainer \cite{chainer_learningsys2015}, and MXNet \cite{mxnet} now also include tools to automatically use multiple GPUs in training neural networks.  Specifically, these tools leverage the fact that SGD and its variants exhibit data parallelism, a simple and powerful motif for scaling \cite{Hillis, blelloch}.  Data parallelism simply requires that a computation can correctly be performed by reducing (or gathering) the results of independent calls over arbitrary data subdivisions, and it has wide applicability to computations amenable to GPU acceleration \cite{Nickolls}.

\subsection{Our Framework}
Although it shares many elements with those examples, our framework offers perhaps a unique blend of generality and ease of use.  It does so by implementing a level of abstraction that addresses the gap left between manual programming of individual devices and pre-fixed multi-GPU training routines.    Synkhronos\footnote{\texttt{https://github.com/astooke/Synkhronos}} automatically coordinates multiple devices to operate as if one, but in a way fully general to any data-parallel computation programmable in Theano.  

Platoon\footnote{\texttt{https://github.com/mila-udem/platoon}} is one previous Theano extension supporting asynchronous multi-GPU computation, wherein workers are programmed separately from the master thread.  In contrast, our framework implements parallel execution automatically at the level of individual Theano functions, which prescribes synchronicity.   This dramatically simplifies user development.  Usage primarily entails a few line-for-line code exchanges from Theano, and the user's programming responsibilities remain entirely within the original (serial) program.

At the same time, our framework is designed with scaling performance in mind.  The multi-processed implementation ensures concurrent operation\footnote{i.e., in the presence of Python's Global Interpreter Lock} for high numbers of devices, without requiring MPI.  Inclusion of a Numpy-like shared memory interface (also shareable with other Python processes) and use of the NVIDIA Collective Communication Library (NCCL) combine to enable nimble management of both GPU and CPU memories.  Lastly, computations are organized to minimize GPU-to-CPU transfers, wherever possible.

\subsection{Paper Organization}
In this paper we describe our framework from a systems perspective, as follows.  First, relevant background on Theano's interface is provided, leading to responsibilities for Synkhronos.  Then, we describe the techniques used for employing multiple devices, including matters of setup, synchronization, and memory management.  We also introduce additional interface features useful for common data parallel compute patterns, including automated input slicing to avoid out-of-memory errors and input indexing for efficient memory sharing and shuffling.  Finally, we investigate performance results in a representative case study before concluding with our future outlook.  Readers already familiar with Theano or other frameworks may wish to view directly the example code comparison provided in Appendix~\ref{app:code}.

\section{Background: Design Requirements}

To formulate the functionalities required of Synkhronos, we first review relevant features of the Theano interface it must interact with: functions and shared variables.

\paragraph{Theano Functions} After constructing symbolic computation expressions using Theano, the user builds functions by specifying the expressions to be computed--and the corresponding input variables--to Theano's ``function'' method.  Theano compiles optimized code (for CPU or GPU), which the user later executes by calling the function on data inputs.  Example functions may compute a forward pass through a neural network, or a gradient over a minibatch of inputs and labels.

\paragraph{Theano Shared Variables} A Theano shared variable has a data array associated with it and can serve as an implicit function input.  Two reasons for using them include: 1) the data persists across different functions, and 2) the data can be held on the GPU, eliminating latency of transmission from the CPU during function calls.  One common use is for the weights of a neural network layer.

\paragraph{Theano Updates} Aside from returning outputs, Theano functions can also modify shared variables, in an operation termed an ``update''.  It is common practice, for example, to build a function which computes a gradient and immediately uses it to update network parameters in place.

\paragraph{Synkhronos Requirements} In sum, Synkhronos must prepare Theano functions for each GPU and make the relevant shared variables available on each device.  Then it must provide means to 1) run functions on all GPUs simultaneously and reduce the results and 2) modify each device's shared variables, including translating local updates into global ones.

\section{Synkhronos Program Flow}
Given what functionality Synkhronos must perform relative to Theano, we now describe our design decisions by way of a program overview.  Data parallelism and synchronicity are two defining characteristics simplifying the design.   Overall functionality is summarized in Figure~\ref{fig:diagram}, and brief example code appears in Appendix~\ref{app:code}.  Discussion on data management is deferred to the following section.

\subsection{Computation Setup}

\paragraph{Fork} Synkhronos automatically forks a separate Python process for each additional GPU. A barrier across all processes guards the start and finish of any user call requiring action by workers, preventing any race conditions.  The user's process acts as master \emph{and} participates as a parallel worker; it remains available for single-GPU use through Theano.

\begin{figure}[t]
  \centering
  \frame{\includegraphics[width=0.9\linewidth]{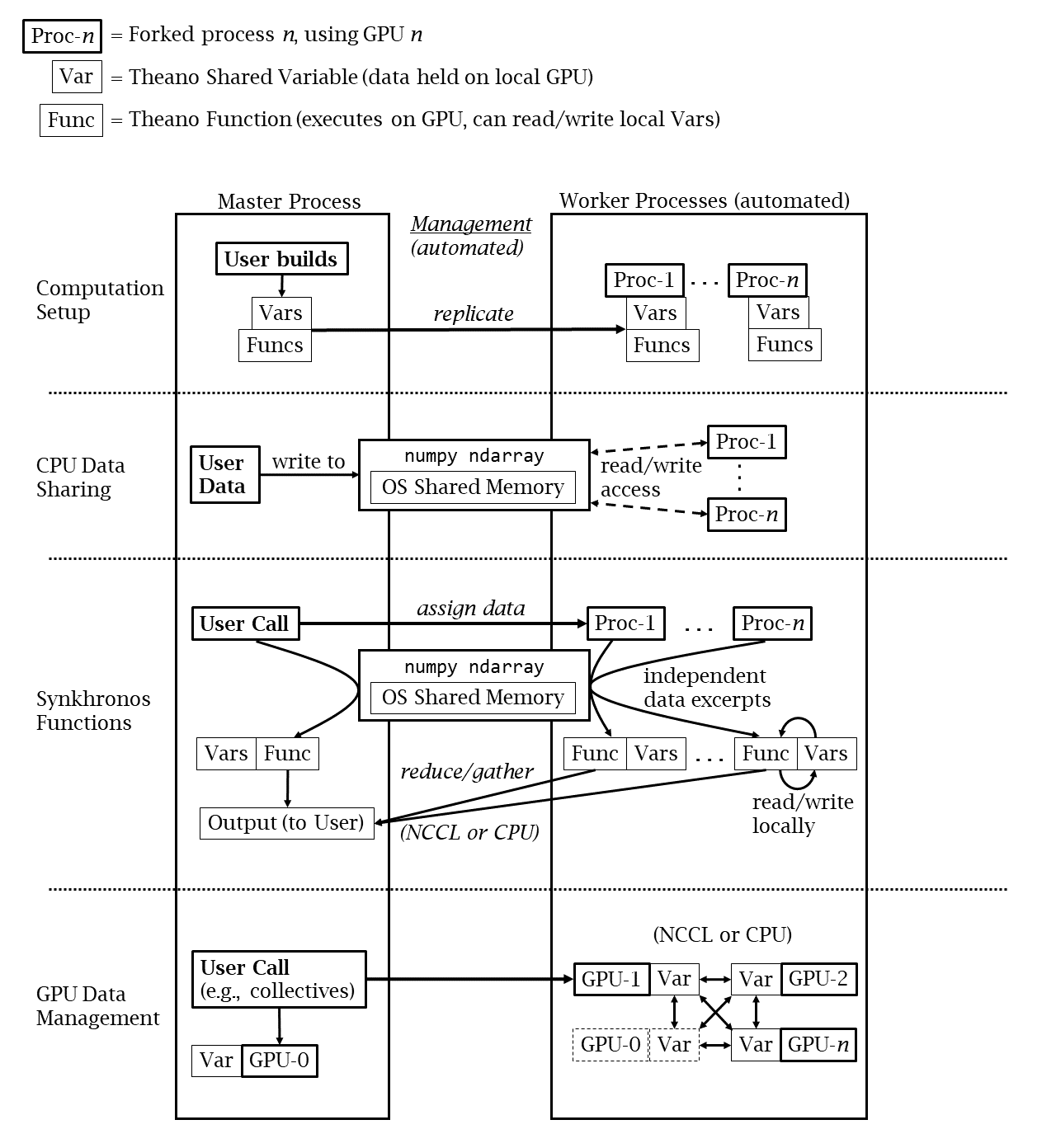}}
  \caption{Synkhronos overview: automated worker management for synchronous computations and communications.}
  \label{fig:diagram}
\end{figure}

\paragraph{Build Symbolic Expressions} After forking, the user builds expressions as usual with Theano (or other extensions, e.g. Lasagne \cite{lasagne} for neural networks), and shared variables are thereby allocated on the master GPU.  This can comprise a majority of the user's overall Theano code--Synkhronos leaves it untouched.  

\paragraph{Build Functions} Synkhronos presents a ``function'' method, replacing Theano's.  Internally, it builds underlying Theano functions.  The main difference to the user is the ability to specify a reduce/gather operation to use for each output.  Thereafter, all functions are distributed to the worker processes using Theano's function serialization, via pickling.  When unpickling, all involved shared variables are automatically replicated on every GPU.

\subsection{Running Computations}

\paragraph{Function Calling} Following distribution, the user calls Synkhronos functions just as Theano functions, but now all GPUs participate in computation.  A function call induces the sequence: 1) data inputs are scattered equally (as possible) across workers, 2) each device calls the same Theano function on its assigned data, and 3) results are reduced or gathered back to the master process and returned to the user.  For program clarity, function updates to shared variables are applied only locally within each GPU.

\paragraph{Shared Variable Management} A new element beyond single-device Theano programs is the management of (replicated) shared variables across multiple memories.  Synkhronos provides several MPI-like colectives, such as \textit{broadcast} and \textit{all-reduce}, which can use NCCL (via the Pygpu package) for high-bandwidth inter-GPU communication (including NVLink).  CPU-based collectives are also included, along with the means to get and set values on any individual GPU.

\subsection{An Example: Synchronous SGD}
\label{sec:SSGD}
Using SGD as an example, the computation sequence could be as follows: 1) a first Synkhronos function takes in data and computes the gradient, storing it in a shared variable (local on each GPU), 2) this variable is all-reduced using Synkhronos, and finally 3) a second Synkhronos function applies the update rule in each GPU using the combined gradient values.  For convenience, such an adaptation of all update rules included in Lasagne is provided , namely momentum \cite{Nesterov}, RMSProp \cite{rmsprop}, and Adam \cite{adam}, among others.  These store the gradients of all variables into one (flattened) array for faster inter-GPU communication.    

\section{Synkhronos Data Management}
Synkhronos must transfer users' data inputs, typically in the form of Numpy arrays, from the user process to the worker processes.  For fast performance, memory copies (within the CPU) and memory transfers (CPU-GPU) should be kept to a minimum.  Synkhronos includes special data objects and communication collectives for this purpose.

\subsection {Synkhronos Data Objects (for Function Inputs)}
Input data is communicated to worker processes using operating system shared memory,\footnote{It is similar to a Multiprocessing RawArray (see section 17.2.2.6.1 at \texttt{https://docs.python.org/3.5/library/multiprocessing.html}, except allocatable before or \emph{after} forking.} exposed to the user in a special data object.  Every process has equal read-write access, so workers can concurrently feed their Theano functions by excerpting their assigned data in parallel.  Synkhronos wraps each shared memory allocation in the Numpy array interface, allowing full array indexing for user reading and writing.  Writing an entire data set (or a large chunk) into such an array obviates the need for any future memory copies, with the use of input indexing (see Section~\ref{sec:indexing}) if needed, e.g. for minibatches.  The lowest tensor dimension is taken to represent independent data points for scattering inputs.\footnote{Under C-style memory layout this provides contiguous memory assignments.}  Inputs designated for broadcast are simply used as is.  

The size of the underlying memory allocation can be made greater than the size of the outward facing Numpy array, preventing reallocation when growing or shrinking an array later.  Special methods are provided for reshaping these arrays and freeing their memory.  Conveniently, the Numpy array sub-object can be passed to other user Python processes to be read/written as shared memory.  In sum, the aim is to present an interface which allows the user to optimize memory performance given the multi-process context.  

\subsection{Scattering to GPUs (for Shared Variables)}
Programs with re-used inputs are often more efficient when data is stored on the GPU over multiple function calls, and the increased aggregate memory of multiple GPUs makes this feasible in a greater number of cases.\footnote{Our original motivation was a big-batch, hessian-free learning algorithm, which achieved super-linear speedup by fitting each batch across multiple GPUs.}  Synkhronos helps with its  ``scatter'' collective, which evenly divides input arrays into shared variable storage across GPUs.  It uses the same scattering scheme as for explicit inputs, i.e. by first tensor dimension.  This is a convenient one-line replacement for setting multiple device memories manually (also possible), and it can make use of Synkhronos data objects and input indexing.

\section{Synkhronos Extensions to Theano Function Interface}
Synkhronos includes two extensions to the Theano interface for calling functions.  These support common data-parallel compute patterns, simplifying user code and improving program efficiency.

\subsection{Automated Input Slicing \& Aggregation}
The first extension is automated input data slicing, which can be used to avoid out-of-memory errors during computations too large to do in one call to a device.  When slicing is used, workers compute their results by aggregating over multiple calls to the underlying Theano function, each using a subset of the worker's assigned data.  Slice results are aggregated in-place on the GPU.  Worker results are reduced once back to the master process. If the function includes updates, all slices are computed using the original values, with updates accumulated and applied only once at the end.  Automated slicing also applies to implicit inputs, i.e. data stored in advance on the GPU in shared variables (the memory requirement during computation is often much larger than the stored data).  The user provides the number of slices to use as an optional input at each function call.

\subsection{(Parallel) Input Indexing}
\label{sec:indexing}
The other extension is input indexing, whereby the user can specify which data elements to use during a function call.  Designated indexes can make a slice or can be a list of (random) indexes.  This is efficient for shuffling, which necessitates a memory copy, as each work can excerpt its own share of the indexes in parallel from the shared memory; no excess memory copies occur.  Similar functionality is provided for inputs stored on-GPU in shared variables, where the same or different indexes can be applied to each device.  Automated input slicing applies internally in each worker, after input indexing has determined assignments.   

\section{Scaling Performance}
We explored an example case study to answer whether Synkhronos can achieve linear scaling on a modern supervised learning problem.  We timed the training of a ResNet-50 \cite{resnet} model using SGD on an NVIDIA DGX-1 (8x Tesla P100 GPUs, NVLink).  Several gigabytes of synthetic images were generated as random single-precision arrays of dimension 3x224x224, and each timing run consisted of the same number of epochs over the entire data set, lasting at least several minutes.  

\begin{figure}[ht]
  \centering
  \includegraphics[width=0.8\linewidth]{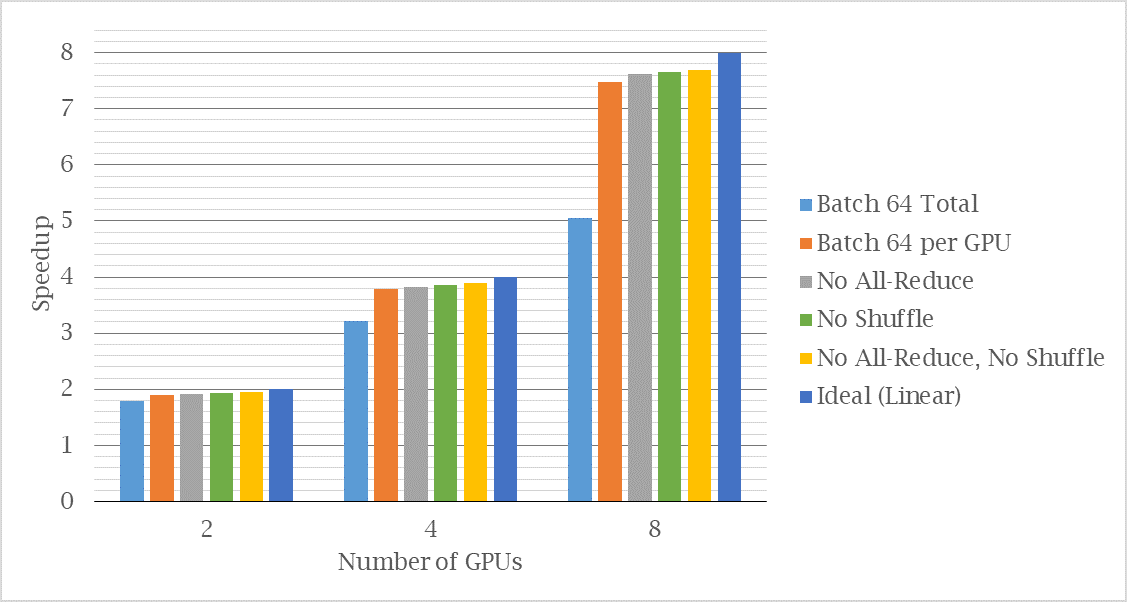}
  \caption{Speedups of ResNet-50 training relative to single-GPU, Theano-only program, with ablations.}
  \label{fig:scaling_bar}
\end{figure}

The base case was a Theano-only program running on a single GPU with the rest of the machine idle, using a batch size of 64 images.  We ran separate base timings for data shuffling or not.  Results from several multi-GPU configurations are shown in Figure~\ref{fig:scaling_bar}, all obtained using Synkhronos.  No input slicing was used, but input indexing was, and all input data was passed as explicit inputs to the function call (i.e., not stored in advance on the GPU).  The network update scheme was as described in Section~\ref{sec:SSGD}.

\subsection{End-to-End Training Timing}

\paragraph{Batch - 64} The first test explored the scaling while keeping the algorithm perfectly unchanged.  The total batch size remained fixed at 64, so the amount of data processed by each GPU in each minibatch scaled down with the number of GPUs.  This yielded sub-linear scaling above 2 GPUs, as GPU utilization decreased, with a speedup only slightly greater than 5x when using all 8 GPUs.  

\begin{figure}[ht]
  \centering
  \includegraphics[width=0.5\linewidth]{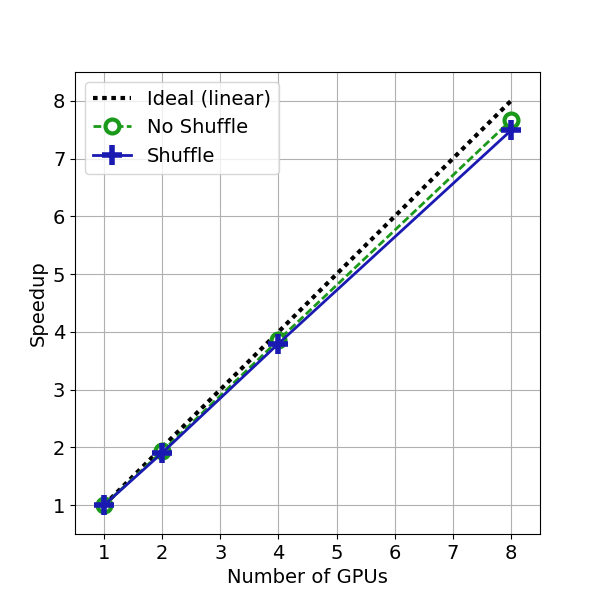}
  \caption{Speedup (near-linear) of ResNet-50 training relative to single-GPU, Theano-only program; with batch scaling.}
  \label{fig:scaling_plot}
\end{figure}

\paragraph{Batch - 64 $\times$ \#GPU}The remainder of the tests used an algorithm modified by scaling up the minibatch size so that each GPU processes 64 images at every call, following the technique successfully applied in \cite{Imagenet_1_Hour} to train to convergence on ImageNet in under one hour.  This improved the scaling of the fully-communicating algorithm up to 7.5x with shuffling and 7.6x without, as compared against their own respective base cases--nearly linear, as seen in Figure~\ref{fig:scaling_plot}.  The base Theano code processed 1.75 minibatches per second, giving training speeds for 1-GPU Theano and 8-GPU Synkhronos of roughly 110 and 830 images per second, respectively.

As shown in Figure~\ref{fig:scaling_bar}, turning off the gradient all-reduction improved scaling only slightly, to 7.6x with shuffling and nearly 7.7x without.  This confirmed a small cost of communication compared to the computation in this scenario (note that Synkhronos permits any intermediate level of communication frequency).

Another view is to measure using the 2-GPU case as the baseline, i.e. as 2x speed, which highlights scaling alone by de-emphasizing any fixed overhead in Synkhronos.  Under this comparison, both the shuffle and non-shuffle algorithms achieved >7.8x speeds using 8 GPUs.  This indicates high likelihood for good scaling up to greater numbers of GPUs.

\subsection{Detailed Profiling}
Lastly, we examined the main use case in more detail in an attempt to understand scaling imperfections.  These tests included full communication, data shuffling, and batch scaling, and they used CUDA launch blocking to enforce CPU-GPU synchronization for proper profiling. Table~\ref{table:profiling} contains a summary of the training loop's main elements and their timings in a short run, comparing Theano-only code against 8-GPU Synkhronos. 

Starting from the bottom of the table, the all-reduce provided through NCCL added less than one percent in overhead.  The exact timing of shuffling and computing naturally varied from call to call, leading to a slight slowdown from the straggler effect, in this case roughly 2\%.  Shuffling scaled well, to 7.9x, leaving some uncertainty as to the source of diminished scaling in the end-to-end tests.  Further profiling gathered through Theano indicated a 1.8\% time overhead for CPU-to-GPU data transfer within the Theano function.  This increased to 3.6\% in the 8 GPU case, as expected on the given hardware, accounting for much of the loss to 7.7x in the Theano function calls.  Overall, we measured near-linear scaling of 7.5x, matching the sum of these components' contributions.  

\begin{table}[ht]
  \caption{Detailed profiling, ResNet-50 SGD training (CUDA launch blocking enabled)}
  \label{table:profiling}
  \centering
  \begin{tabular}{lccc}
  \toprule
   & \multicolumn{2}{c}{Time (s)} \\
  \cmidrule{2-3}
  Item & Theano & Synk-8 & Scaling \\
  \midrule
  Total & 395.1 & 52.6 & 7.5 \smallskip\\
  Theano Function & 382.5 & 49.5 & 7.7 \\
  Shuffle & 12.6 & 1.6 & 7.9 \\
  Straggler Effect & -- & 1.0 & -- \\
  All-Reduce Gradient & -- & 0.46 & -- \\
  \bottomrule
  \end{tabular}
\end{table}

\section{Conclusion}
We have presented Synkhronos, an extension to Theano for computing with multiple devices under data parallelism.  After detailing the framework and functionality, we demonstrated near-linear speedup on a relevant deep learning example, training ResNet-50 with 8 GPUs on a DGX-1.  The design emphasizes easy migration from single- to multi-device programming by the user while maintaining full generality to any data parallel computation.  It includes flexible tools for efficient memory management, yet currently remains limited to single-node computing.  Lastly, because it is written entirely in Python, the package is more widely accessible to modification by interested users.  We refer to the code repository\footnote{\texttt{https://github.com/astooke/Synkhronos}} for further examples.  We hope that this package will accelerate the work of researchers and developers who use it, and that it may contribute helpful concepts for multi-device interfaces for other performance computing frameworks going forward.

\subsubsection*{Acknowledgments}
We thank Fr\'{e}d\'{e}ric Bastien, Pascal Lamblin, and Arnaud Bergeron for discussions and pointers throughout development, and the entire Theano development team for their base of work.  Thanks again to Fr\'{e}d\'{e}ric and to Carlos Florensa, Peter (Xi) Chen, and Anushree Saxena for helpful comments on the manuscript. The DGX-1 used for this research was donated by the NVIDIA Corporation.  Adam Stooke gratefully acknowledges the support the Fannie and John Hertz Foundation.

\bibliographystyle{unsrtnat}
\bibliography{main.bbl}

\newpage
\appendix

\section{Code Example}
\label{app:code}
In the figures below we present a simple code example for running stochastic gradient descent in Theano, which is revised into Synkhronos.  In the Theano example, the functions \verb+build_cnn()+ and \verb+setup_training()+ will contain other Theano code (or possibly Lasagne), which is left untouched in the Synkhronos program.  The \verb+fork()+ command automatically uses all GPUs if not otherwise specified, and the \verb+distribute()+ command replicates all functions and Theano shared variables on the worker GPUs.  Remaining line numbers where Synkhronos is active are highlighted.  The \verb+data()+ command is used here to write the training data to operating system shared memory (once).  The main change in the training loop is the call to all-reduce the network parameters, which in the case of simple SGD preserves the algorithm.  Further examples and demonstrations can be found at the code repository.    

\begin{figure}[ht]
  \centering
  \frame{\includegraphics[width=0.9\linewidth]{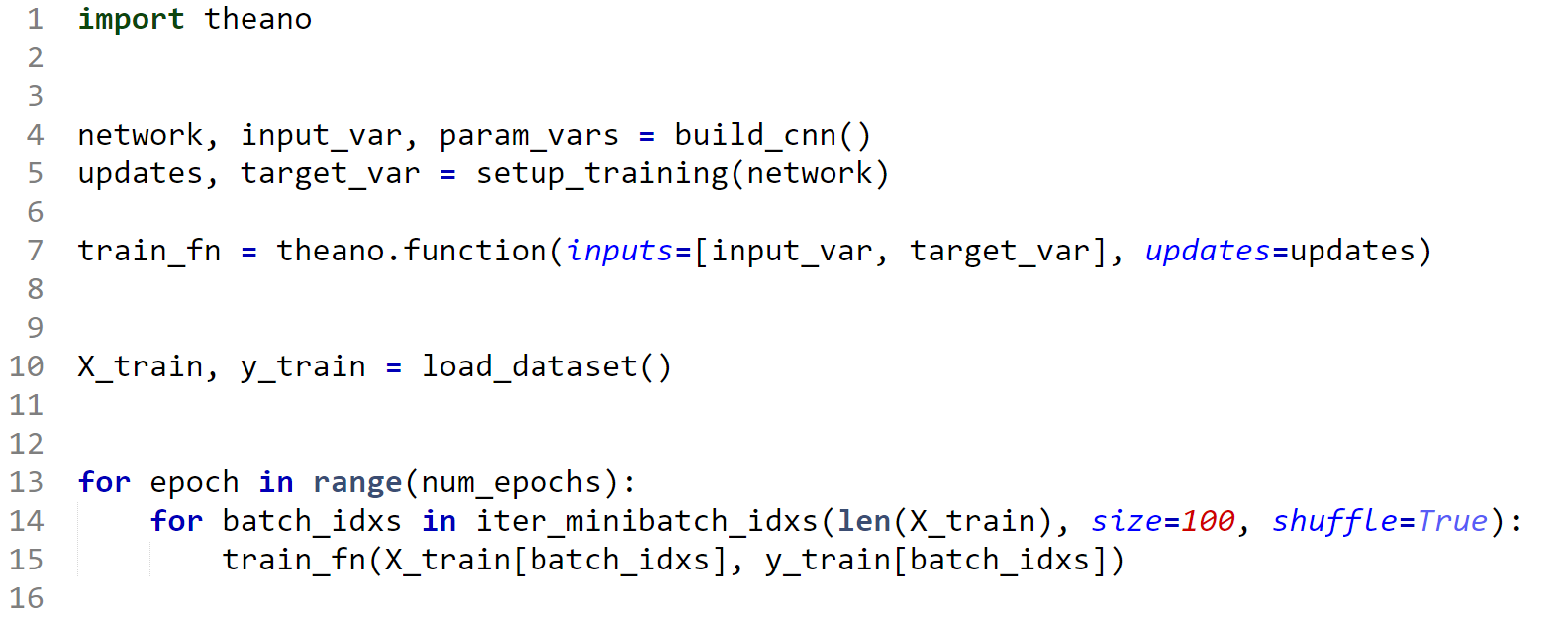}}
  \caption{Theano program for single-GPU SGD.}
  \label{fig:code_theano}
\end{figure}

\begin{figure}[ht]
  \centering
  \frame{\includegraphics[width=0.9\linewidth]{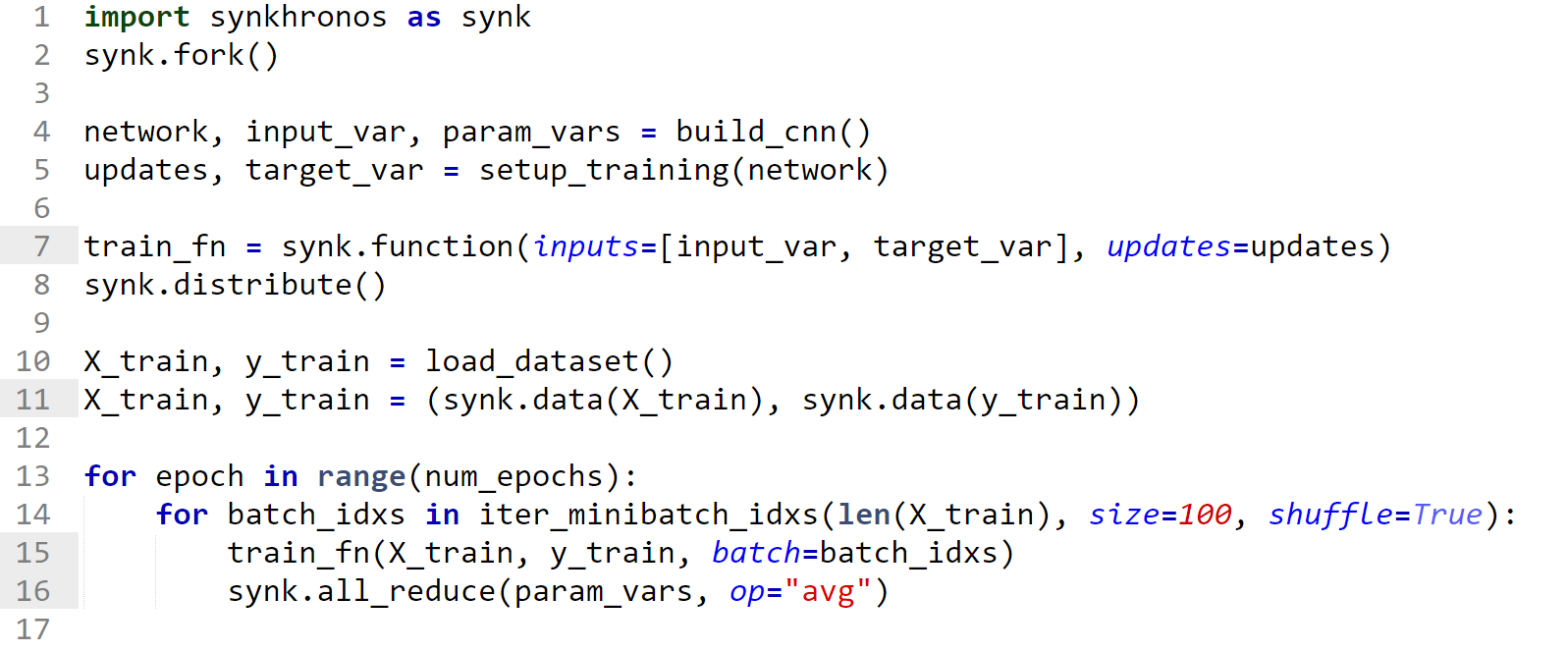}}
  \caption{Synkhronos program for multi-GPU SGD.}
  \label{fig:code_synk}
\end{figure}

\end{document}